\documentclass{appolb}
\usepackage{epsfig}

\begin{document}
\title{Introduction to
Low $x$ Physics
and Saturation
}
\author{N. Armesto
\address{Theory Division, CERN\\ CH-1211 Gen\`eve 23, Switzerland}
}
\maketitle
\begin{abstract}
The idea of saturation of parton densities in small $x$ physics
is briefly introduced. Some
aspects of saturation are described, mainly focusing on the status of our
knowledge on the non-linear equations describing the high parton density
regime.
Implications of saturation ideas on the description of nuclear collisions
at the Relativistic Heavy Ion Collider are discussed.
\end{abstract}
  
\section{Introduction: small $x$ and saturation}
The BFKL equation \cite{bfkl} resums gluon
ladders taking into account all leading contributions $[\bar \alpha_s
\ln{(1/x)}]^n$ (LL$1/x$), with $\bar \alpha_s=\alpha_s N_c/\pi$ and
$x$ the fraction of momentum of the hadron carried by the parton.
Although it was
originally an attempt to compute the high-energy asymptotics of QCD and to
justify Regge Theory, it turned out to be an evolution equation in
$1/x$ for the so-called unintegrated parton distributions
(used in $k_T$-factorization \cite{ktfac}
to compute inclusive particle production at
scale $Q$, $\Lambda_{\rm QCD} \ll Q \ll E_{cm}$), which behave
$\propto x^{-\bar \alpha_s 4 \ln 2}$.
Experimentally \cite{exphere} $xG(x,Q^2) \propto x^{-\lambda}$, with
$\lambda \sim 0.3$ sizeably smaller than predicted by BFKL for
reasonable values of $\alpha_s$.

Both $k_T$-factorization and the BFKL equation are valid in a low parton
density, linear regime.
At small $x$ (equivalent to large energies for a fixed scale) parton
densities become high. Then the idea of saturation
of parton densities
(see
\cite{cargese}) becomes unavoidable, as parton fusion balances
parton splitting if $A_{\mu \nu} \propto 1/g$. It can be alternatively
formulated in terms of the $S$-matrix at fixed impact parameter $b$
becoming black, $|S(b)|=0$.
Saturation constitutes a new regime of QCD, in which old ideas (pomeron
interaction, multiple scattering,
coherence arguments,$\dots$) are expressed in a new, QCD language.
It also offers a link between small $x$ physics and heavy ion
collisions: in nuclear collisions at high energies large partonic
densities are expected, both due to the high energy and to the
$A^{1/3}$-enhancement coming from the nuclear size. The understanding of the
initial state in a nuclear collision is crucial to get a coherent
picture of the eventual thermalization of the system and creation of
Quark Gluon Plasma.

\begin{figure}[htb]
\begin{center}
\epsfig{file=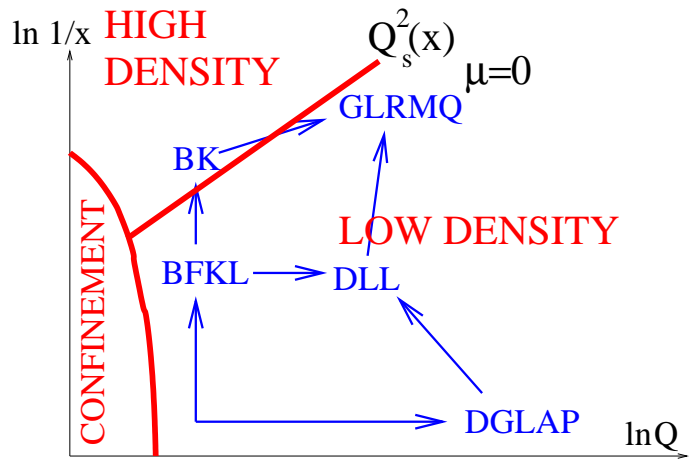,width=9cm}
\end{center}
\caption{Regimes of QCD in the $\ln{(1/x)}$-$\ln{Q}$ plane.}
\label{fig1}
\end{figure}

Our present understanding of the
different regimes of QCD is summarized in Fig.~\ref{fig1}. At low $Q^2$ we
have the confinement region. At large $Q^2$ and not too high $1/x$ we
are in the low-density region, where the usual evolution equations can be used:
BFKL in $\ln{(1/x)}$ and DGLAP in $\ln{Q^2}$. For both large
$\ln{(1/x)}$ and $\ln{Q^2}$ we are in the DLL regime,
where a first non-linear
correction has been proposed, the GLRMQ equations
\cite{glrmq}.
Finally in the high-density regime,
separated from the low-density one by
a line which defines the saturation scale $Q_s(x)$, a non-linear
generalization of
BFKL is the Balitsky-Kovchegov (BK) equation
\cite{bk1,bk2}.

I will present in the next Section the phenomenological
models implementing saturation and our current theoretical understanding
of this new regime. In
Section
3 I will analyze properties of the solutions of the BK equations, and in
Section 4
the application of saturation ideas to high-energy nuclear collisions,
presently studied at
the Relativistic Heavy Ion Collider (RHIC) at BNL. Finally
I will draw some conclusions.
Due to space limitations, many interesting topics
like NLL developments\footnote{NLL effects are expected to be
important at larger rapidities, $Y> 1/\alpha^{5/3}_s$,
than unitarity corrections, important for
$Y> \ln{(1/\alpha_s)}/\alpha_s$.},
non-linear equations
in the collinear approach, relations with other realizations of
collectivity
or many examples of observables analyzed, will not be
discussed (see e.g. \cite{cargese,pajares,pahp} and references therein).

\section{Present realizations of saturation}

\subsection{Phenomenological models}
The most compelling indirect evidence of saturation comes from the
phenomenological success of some models containing saturation ideas when
confronted to HERA data. The most commonly used is the
GBW
model \cite{gbw}. It is formulated within the color dipole model,
in which the interaction of the virtual photon with the hadron or nucleus is
described as
a convolution of the probability that the photon fluctuates into a
$q\bar q$ pair of fixed transverse size $r$ with the dipole-hadron cross
section\footnote{The fluctuation length of the photon into the $q\bar q$,
$\propto 1/x$, becomes at small $x$ larger than the hadron size, so its
interaction is coherent.}.
For the latter, the GBW model provides an ansatz for the scattering amplitude
$N$:
\begin{equation}
{\sigma^{q \bar q-h}(r,b)\over 2}=
N(r)=1-e^{-Q_s^2r^2/4},\ \ 2\int d^2b \equiv \sigma_0,
\ \ Q_s^2=\left ({x_0 \over x}
\right )^{\lambda}\ ,
\end{equation}
which implements the unitarity limit,
$\sigma^{q \bar q-h}(r,b)=2\ {\rm Re}[1-S(r,b)]\leq 2$, in a very simple way.
With $\sigma_0\sim 20$ mb, $x_0 \sim 3\cdot 10^{-4}$ and $\lambda \sim 0.3$,
this model gives a reasonable description of all HERA data
on $F_2$ of the
proton for $x<0.01$ ($Q^2< 450$ GeV$^2$). It has been extended to
included DGLAP evolution and to describe diffraction at HERA \cite{gbhere}, and
is widely used for phenomenology \cite{szhere}.
It implies a
$[\tau=Q^2/Q^2_s(x)]$-scaling \cite{scal} observed at HERA and
also searched in nuclear data \cite{nucscal},
but it is unclear if the expectation
$Q_{sA}^2 \propto AT_A(b) \propto A^{1/3\div 2/3}$ \cite{cargese}
is fulfilled at present 
$x$, $Q^2$ \cite{mio}.

\subsection{Theoretical developments}
After earlier studies \cite{glrmq,earlier},
a milestone in the theoretical development of saturation ideas in QCD was
the MV model \cite{mv}. This model
treats classical radiation
from color sources
moving ultrarelativistically through a large nucleus.
With a form for the color correlators in the
target and taking into account the non-abelian gluon interaction, it gives an
explicit form for the nuclear gluon distribution in the transverse phase space,
\begin{equation}
{dN_g^A \over d^2b d^2l}\propto \int {d^2r \over r^2}\ e^{-il\cdot r}
[1-e^{-Q_s^2r^2\ln{[(\Lambda_{\rm QCD} r)^{-2}]}/4}].
\end{equation}

Later on, gluon radiation (quantum evolution)
of color sources was introduced,
leading to a renormalization group-type equation (the so-called Color Glass
Condensate \cite{cgc}).
The rescattering of the projectile in the
nucleus is described through Wilson lines whose average on
target configurations
gives the $S$-matrix. The key point is that
the fields (occupation numbers)
become large ($F_{\mu \nu}, f_g\propto 1/g^2$)
but the coupling is
small, so 
classical arguments and perturbative methods are applicable. A hierarchy of
coupled
equations for $n$-gluon correlators appears \cite{bk1,cgc}.

\section{The Balitsky-Kovchegov equation}
In the framework outlined in the previous Section,
the BK equation \cite{bk1,bk2} is the evolution equation in $1/x$ for the
2-gluon correlator, decoupled from the hierarchy in the
$N_c \to \infty$ limit (and with correlations neglected
\cite{ianmue})\footnote{It was
deduced
in the color dipole model \cite{bk2}, and also (to my knowledge) in an
eikonal approach
\cite{kw} and in BFKL summing fan diagrams in the large $N_c$
limit \cite{misha}.}. As BFKL, it is infrared stable, mixes all twists and
$\alpha_s$ is fixed. For $N(x_1,x_2)\equiv
N_{x_1x_2}$ given
by a target average of the Wilson lines of a $q$ and a $\bar q$
located at transverse positions $x_1$ and $x_2$ respectively, it reads
\begin{equation}
\frac{\partial N_{x_1x_2}}{\partial Y}=\bar \alpha_s
\int \frac{d^2z}{2 \pi} \frac{(x_1-x_2)^2}{(x_1-z)^2(z-x_2)^2}
[N_{x_1z}+N_{zx_2}-N_{x_1x_2}-N_{x_1z}N_{zx_2}],
\label{eq3}
\end{equation}
with $Y=\ln{(x_{initial}/x)}$. Taking into account just the linear terms this
equation reduces to BFKL. In the DLL limit, GLRMQ is recovered \cite{bk2}.
(\ref{eq3}) is the most simple tool we have at our disposal to study the
high-density, non-linear regime. Its analytical solutions are unknown. Only
some theoretical estimates exist
\cite{anal1,anal2,pesch}. So, numerical methods
have been developed.

\begin{figure}[htb]
\vskip -0.3cm
\begin{center}
\epsfig{file=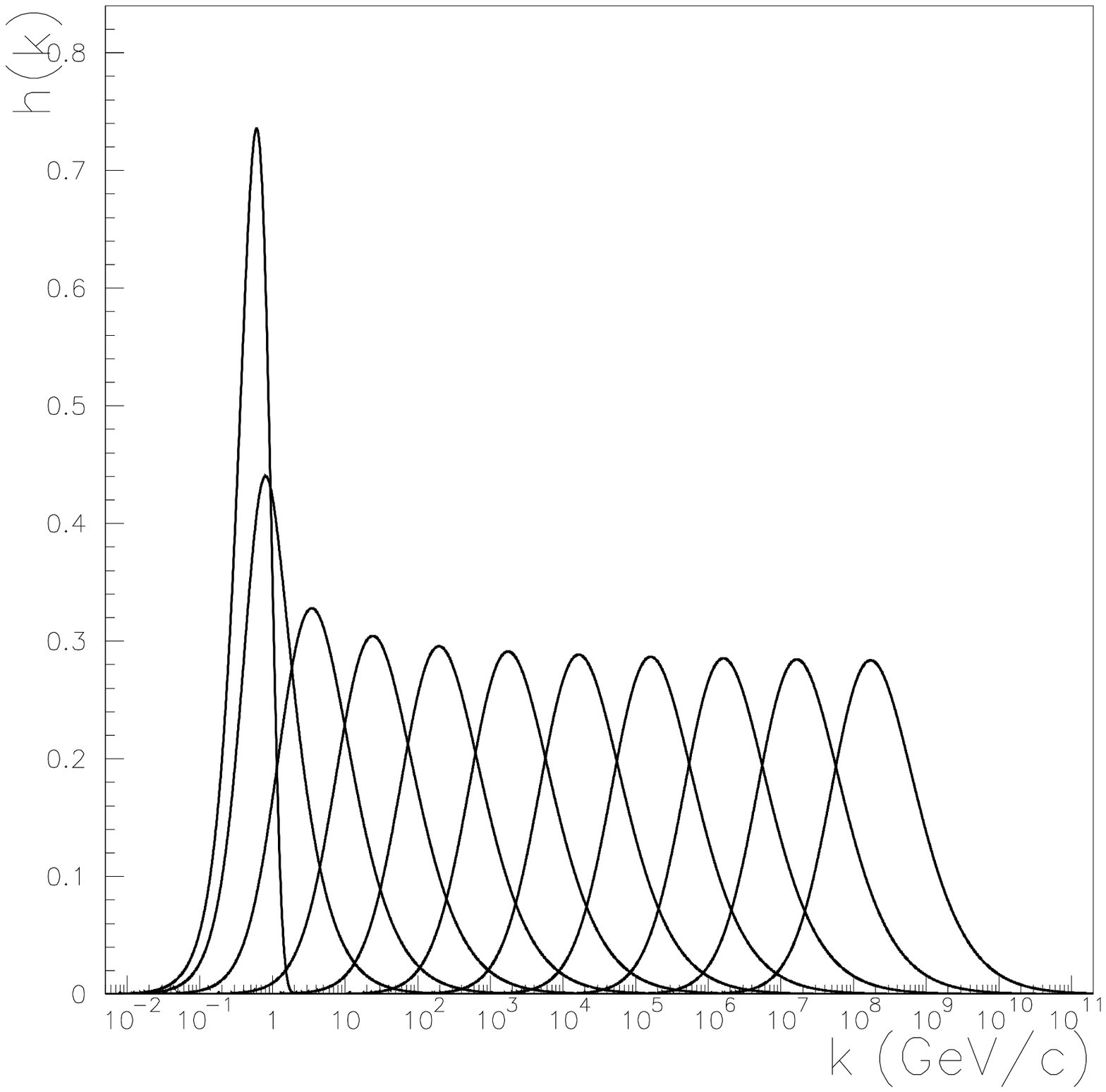,width=6.5cm}\epsfig{file=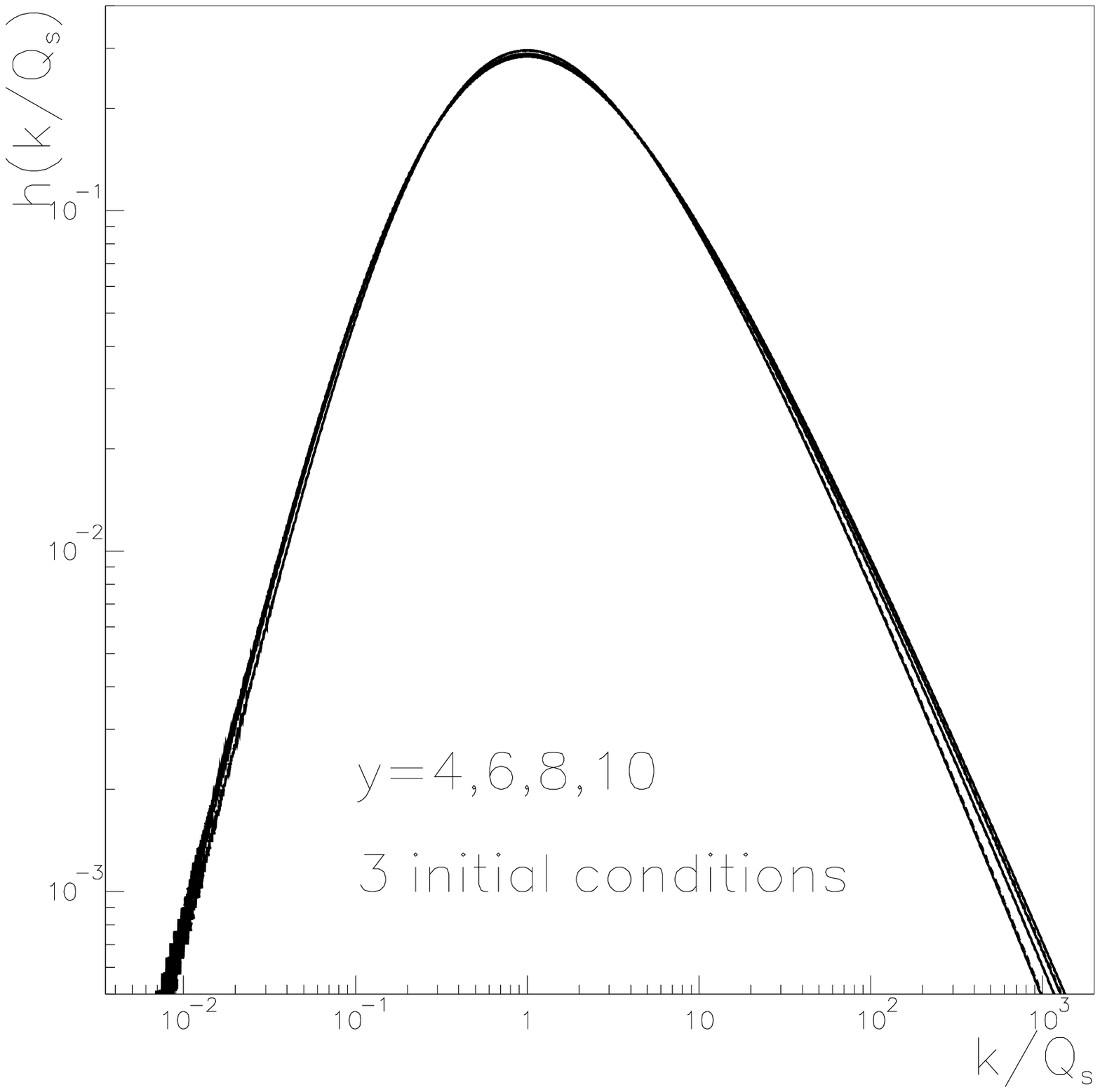,width=6.5cm}
\end{center}
\vskip -0.4cm
\caption{{\it Left}: function $h$ versus $k$ for different $y=\bar \alpha_s Y=
0,1,2,\dots ,10$. {\it Right}: the same function for $y=4,6,8,10$ and three
different initial conditions, versus $k/Q_s$. See
\protect{\cite{albacete}} for further explanations.}
\label{fig3}
\vskip -0.5cm
\end{figure}

These numerical methods usually deal with the situation
$|b|=|x+y|/2\gg |r|=|x-y|$
\footnote{The
solution with full $b$ dependence has also
been studied numerically \cite{stasto}
and
turns out to be of great importance to compute
the behavior of the total cross section at
very high energies in this approach.},
either in position space \cite{levin,lublinsky} or in
momentum space \cite{misha,gbms,ours,albacete} (in this latter case we define
$\phi(k)=\int \frac{d^2r}{2\pi r^2}
\ \exp{(ik\cdot r)}
N(r)$). Using different techniques two
very interesting properties of the solutions
have been found. First, function $h(k)=k^2\nabla_k^2\phi(k)$ gets a
constant shape in $\ln k$ at large $Y$, moving to the right with constant
velocity \cite{misha}. Second, identifying the position of the
maximum of $h$ with the saturation scale $Q_s$,
the solutions show scaling, i.e. $\phi(k)
=\phi(k/Q_s)$
\cite{lublinsky,ours}
(linked to that in $\tau$ discussed in the previous Section,
sometimes called geometrical scaling).
These two features are illustrated in Fig.~\ref{fig3}.
The velocity of the solution
has been
computed,
$Q_s^2 \propto e^{\bar \alpha_s \bar \lambda Y}$, with
$\bar \lambda\simeq 4.1\div 4.6$ \cite{misha,lublinsky,gbms,ours,albacete}.
As a consequence of these features,
the contribution from
the low-$k$ region is reduced with increasing $Y$,
thus offering a possibility to avoid the infrared problem of BFKL.
These features are independent of the initial condition
(MV \cite{albacete}, BFKL-like \cite{misha,ours}, DGLAP-like
\cite{lublinsky}, Gaussian \cite{gbms} or GBW \cite{ours,albacete}). Finally,
the shape of the solutions above $Q_s$ has been examined
\cite{albacete},
favoring a log-corrected shape \cite{anal2} over a pure power in $k$
\cite{anal1}.

\section{Phenomenology at RHIC}

\subsection{Multiplicities and pseudorapidity distributions}
For midrapidity at $\sqrt{s_{NN}}=200$ GeV,
$0.2<m_T/{\rm GeV}<10$ means $0.002<x<0.1$, so $x$ at RHIC
may well be too high
to apply safely
saturation ideas.
Assuming that $x$ is small enough, which we will do in the following,
multiplicities and their evolution with centrality and
pseudorapidity can be computed in saturation,
usually within a factorized approach
\begin{equation}
{dN_{AB}\over dyd{\bf p}_T\, d^2b}
  \propto {1\over p_T^2}
    \int d{\bf k}_T\, f^{Q_A}(y,k_T)\,f^{Q_B} (y,{\bf p}_T-{\bf k}_T),
\label{facto}
\end{equation}
with $Q_{A,B}$ the saturation momentum of hadrons $A$ and $B$ at some given
centrality, and $p_T$
the transverse momentum. It is still unclear to what extent
this factorized
ansatz is exact and which one is the function $f$ which should
be introduced into this equation \cite{facto}; other approaches in the
semiclassical framework have also been essayed \cite{raju}.
With some ansatz for $f$,
a simple formula is derived \cite{simple},
 ${dN \over dy}\propto s^{\lambda/2} N_{part} e^{-\lambda |y|}
\ln{[Q_s^2/\Lambda_{\rm QCD}^2]}$, $Q_s^2\sim A^{1/3}$,
$\lambda \sim 0.3$ (GBW model). This formula
successfully reproduces multiplicities, their evolution
with centrality and pseudorapidity distributions at RHIC.
See that the deviation from the scaling with the number of participants
$N_{part}$ is due just to the log (coming from a factor
$1/\alpha_s(Q_s^2)$), at variance with what is found in
other approaches \cite{dolo}.
So the question arises
whether RHIC data can be
explained by initial state effects\footnote{Elliptic flow in saturation
models has also been analyzed \cite{pajares,elliptic}.} \cite{steinberger}.

\subsection{Transverse momentum distributions}
The transverse momentum distribution of partons and particles is
expected to be suppressed by the presence of a medium,
the so-called jet quenching. This is usually quantified through
the ratios
\begin{equation}
R_{pA}={ {dN_{pA}\over dyd^2p_T\, d^2b}\over N_{coll}
\,     {dN_{pp}\over dyd^2p_T\, d^2b}}\, , \ \ 
 R_{AA} = { {dN_{AA}\over dyd^2p_T\, d^2b}\over N_{coll}
 \,           {dN_{pp}\over dyd^2p_T\, d^2b}}\, .
\label{ratios}
\end{equation}
Normalized in this way, this ratio goes to 1 at large $p_T$
according to the
usual QCD factorization expectation,
with $N_{coll}$ the
number of binary nucleon-nucleon collisions.
Indeed, such suppression has been
observed in AuAu collisions at RHIC \cite{steinberger}.
Contrary to the jet quenching interpretation,
it has been argued in \cite{klm} that such suppression can be explained by
initial state, saturation effects, so it
should also be present in dAu collisions. These collisions have been
studied at RHIC and an enhancement has been found \cite{dau},
the so-called Cronin effect measured long ago. This leads
to the conclusion that the depletion in AuAu is due
to final state effects.

\begin{figure}[htb]
\vskip -0.8cm
\begin{center}
\epsfig{file=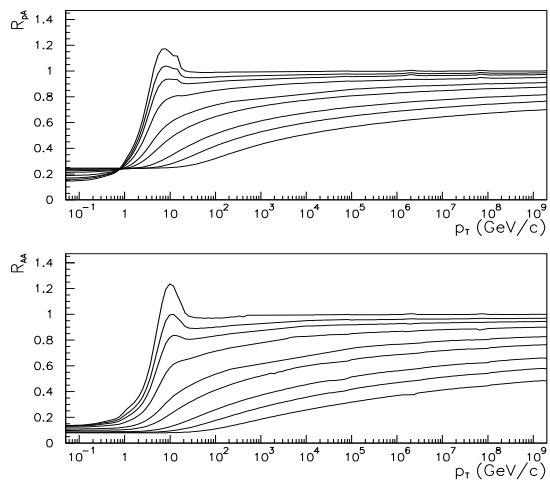,width=8cm,height=7.2cm}
\end{center}
\vskip -0.5cm
\caption{Ratios (\protect{\ref{ratios}}) in pA (upper plot) and AA
(lower plot).
In each plot, lines from top to bottom correspond to rapidities
$y=\bar \alpha_s Y=0$,0.05,0.1,0.2,0.4,0.6,1.0,1.4, 2.0.
See
\protect{\cite{albacete}} for further explanations.}
\label{fig4}
\end{figure}

Let us examine in more detail the result in \cite{klm}. It has been shown
\cite{bkw} that the behavior of $f$ above $Q_s$ is crucial
to get either suppression or enhancement in the ratios
(\ref{ratios}).
The naive expectation is that
saturation effects are important only for $k<Q_s$. Still,
the behavior of $f$ in the region $k>Q_s$ is driven by evolution
\cite{anal1,anal2,albacete}.
Non-evolved
forms lead generally to enhancement \cite{bkw,kkt,jnv}.
On the contrary, in \cite{klm} a form which contains
evolution features \cite{anal1} has been used.
After numerical
studies \cite{albacete}, it has become clear \cite{albacete,bkw,kkt,jnv}
that quantum evolution does
not generate enhancement but very efficiently erases any
that may be present in the initial conditions
(see Fig.~\ref{fig4}). Then a prediction
is that the Cronin effect
will disappear at higher energies (LHC) or for forward rapidities in pA,
corresponding to smaller $x$ in the nucleus. Preliminary BRAHMS data
\cite{lately}
in dAu collisions at $\eta \sim 3$ suggest
such effect. Nevertheless, other effects like running coupling \cite{gbms,rc}
may be important for a
quantitative comparison.

Summarizing, the concept of saturation in small $x$ physics has been
introduced.
Some features of the solutions of the non-linear BK equation which arises in
this context, have been analyzed. The relevance of saturation for the initial
stage of a nuclear collision has been discussed. Assuming that $x$ at RHIC
is small enough to apply saturation ideas,
the importance
of non-linear small $x$ evolution for the interpretation of enhancement or
suppression of the
$p_T$ distributions measured there has been shown.
\vskip 0.1cm
{\small 
\noindent {\bf Acknowledgments:}
It is a pleasure to acknowledge fruitful collaborations
with J.~L.~Albacete,
M.~A.~Braun, A.~Capella,
E.~G.~Ferreiro, A.~B.~Kaidalov, A.~Kovner, C.~Pajares, C.~A.~Salgado and
U.~A.~Wiedemann, and
useful discussions with R.~Baier, D.~Kharzeev, B.~Z.~Kopeliovich,
Y.~Kovchegov,
E.~Iancu, L.~McLerran, K.~Rummukainen, K.~Tuchin,
R.~Venugopalan and H.~Weigert.
I thank C.~A.~Salgado for a critical reading of the manuscript and
the organizers for their invitation to such a nice
meeting. I dedicate this work to the memory of Jan
Kwieci\'nski.
}

\end{document}